\documentclass[a4paper,11pt]{article}
\usepackage{jcappub}
\usepackage{amsmath}
\usepackage{graphicx}

\makeatletter
\gdef\@fpheader{}
\makeatother

\DeclareMathOperator{\trace}{Tr}

\newcommand{\be}{\begin{equation}}
\newcommand{\ee}{\end{equation}}
\newcommand{\bea}{\begin{eqnarray}}
\newcommand{\eea}{\end{eqnarray}}
\newcommand{\barr}{\begin{array}}
\newcommand{\earr}{\end{array}}

\newcommand{\boldmathsymbol}[1]{\ensuremath{\boldsymbol{#1}}}
\newcommand{\sss}[1]{{\scriptscriptstyle{#1}}}
\newcommand{\fiducial}[1]{\hat{#1}}
\newcommand{\like}[2]{\mathcal{L}\!\left(#1 | #2 \right)}

\newcommand{\ur}{\mathrm{r}}
\newcommand{\ub}{\mathrm{b}}
\newcommand{\uf}{\mathrm{f}}
\newcommand{\udm}{\mathrm{dm}}

\newcommand{\unoise}{\mathrm{noise}}
\newcommand{\uS}{\mathrm{S}}
\newcommand{\uK}{\mathrm{K}}
\newcommand{\uA}{\mathrm{A}}

\newcommand{\calP}{\mathcal{P}}

\newcommand{\Mpc}{\mathrm{Mpc}}
\newcommand{\Gpc}{\mathrm{Gpc}}
\newcommand{\muK}{\mu\uK}

\newcommand{\Hstar}{H_*}
\newcommand{\ms}{m_\sigma}
\newcommand{\kr}{k_\ur}
\newcommand{\kf}{k_\uf}
\newcommand{\ellf}{\ell_\uf}
\newcommand{\OmegaB}{\Omega_\ub}
\newcommand{\OmegaDM}{\Omega_\udm}
\newcommand{\ns}{n_\sss{\uS}}
\newcommand{\As}{A_\sss{\uS}}
\newcommand{\Aw}{A_\omega}
\newcommand{\dang}{d_\uA}

\newcommand{\Db}{\boldmathsymbol{D}}
\newcommand{\Cb}{\boldmathsymbol{C}}
\newcommand{\fidlambda}{\fiducial{\lambda}}
\newcommand{\fidCb}{\fiducial{\Cb}}
\newcommand{\fidDb}{\fiducial{\Db}}

\newcommand{\Cnoise}{C_{\unoise}}

\newcommand{\CAMB}{\texttt{CAMB}}
\newcommand{\COSMOMC}{\texttt{CosmoMC}}

\title{Searching for Standard Clocks in the Primordial Universe}

\author[a]{Xingang Chen}
\author[b]{and Christophe Ringeval}

\affiliation[a]{Centre for Theoretical Cosmology,
Department of Applied Mathematics and Theoretical Physics,
University of Cambridge, Cambridge CB3 0WA, United Kingdom}

\affiliation[b]{Centre for Cosmology, Particle Physics and Phenomenology,
  Institute of Mathematics and Physics, Louvain University, 2 Chemin
  du Cyclotron, 1348 Louvain-la-Neuve, Belgium}

\emailAdd{x.chen@damtp.cam.ac.uk}
\emailAdd{christophe.ringeval@uclouvain.be}

\abstract{Classically oscillating massive fields can be used as
  ``standard clocks" in the primordial universe. They generate
  features in primordial density perturbations that directly record
  the scale factor evolution $a(t)$. Detecting and measuring these
  ``fingerprint" signals is challenging but would provide a direct
  evidence for a specific primordial universe paradigm. In this paper,
  such a search is performed for the power spectrum of the Cosmic
  Microwave Background (CMB) anisotropies using the WMAP7
  data. Although a good fit to the data privileges a scale around
  $k=0.01 \,\Mpc^{-1}$, we do not find statistical significance for,
  neither against, the presence of any feature. We then
  forecast the expected constraints a Planck-like CMB experiment can
  impose on the fingerprint parameters by using
  Markov-Chain-Monte-Carlo (MCMC) methods on mock data. We exhibit a
  high sensitivity zone for wavenumbers ranging from $0.01\,\Mpc^{-1}$
  to $0.1\,\Mpc^{-1}$ in which fingerprints show up first on the
  posterior probability distribution of the wavenumber at which they
  occur, and then on the modulation frequency. Within the sensitivity
  zone, we show that the inflationary paradigm can be inferred from a
  single feature generating at least a $20\%$ modulation of the
  primordial power spectrum. This minimal value sensitively depends on
  the modulation frequency.}


\begin{document}

\maketitle

\section{Introduction}

Experimentally distinguishing the primordial universe paradigms that
lead to the Big Bang model is an outstanding challenge in modern
astrophysics and cosmology. The leading candidate is inflation
\cite{Guth:1980zm, Linde:1981mu, Albrecht:1982wi}. While we are still
gathering experimental evidences to distinguish the inflationary paradigm from
other possible alternatives, within the last 15 years we unexpectedly
discovered that our late-time universe is actually inflating. One
important reason this discovery is so convincing is that we are able
to directly measure the scale factor of the universe $a$ as a function
of time $t$. Using the type Ia supernovae as ``standard
candles"~\cite{Perlmutter:1997zf,Riess:1998cb} the measurement of the
magnitudes versus the redshifts of the stars directly tells us $a(t)$
-- the definition of the evolutionary paradigm of the universe.  In
contrast, the information that we have obtained so far from the
primordial density perturbations, such as the approximate
scale-invariant power spectrum, are convoluted consequences of the
scale factor evolution, and this is a primary reason for possible
degeneracies. So, can we directly measure the scale factor as a
function of time for the primordial universe?

It has been recently proposed that we may look for ``standard clocks"
\cite{Chen:2011zf,Chen:2011tu}. Such clocks should have a known
time-dependence and leave their ``ticks" in terms of features in the
primordial density perturbations. They should exist as general as
possible in all paradigms and leave identifiable characteristics in
the density perturbations. Good candidates are classically vibrating
massive fields.

By massive, we mean the masses of these fields are much larger than
the event-horizon energy-scale\footnote{For non-inflationary
  cases, the event horizon energy-scale may not be close to the Hubble
  scale.} during the primordial epoch. Such fields are abundant in any
primordial universe models, for instance in terms of stabilized
moduli. The low energy effective field trajectory, driving the
evolution of the universe, is running in the valleys determined by
these massive fields. All these fields span a multi-field space with
very large dimensions. Generically, one expects the low energy
trajectory to turn from time to time in this multi-field space, and,
depending on the sharpness of the turns, some massive fields
orthogonal to the adiabatic field trajectory may get excited and
oscillate for a while. Such processes have variety of manifestations
in the low energy theory, appearing as turning, sharp features,
particle interactions and etc. The induced oscillations typically have
small amplitudes. For most purposes, they can be safely averaged out
or treated as some small side-effects. However, these side-effects
contain very valuable information. How the massive fields oscillate in
a time-dependent background can be computed precisely and have several
very distinguishable features. These oscillations generate
``cosmological ticks" that can be used as the above mentioned standard
clocks.

The next questions are, how large observational effects can be induced
by these small oscillations, and how model-independently can we make
theoretical predictions? It is shown in \cite{Chen:2011zf,Chen:2011tu}
that three universal properties nicely fit into each other for our
purpose. Firstly, these oscillations imprint standard clocks in
various cosmological parameters in terms of small oscillating
components; and these parameters appear as couplings in the
correlation functions. Secondly, these oscillations affect the density
perturbations through the universal Bunch-Davies vacuum of the quantum
fluctuations, instead of their highly model-dependent event-horizon
scale and super-event-horizon evolution; and this makes general
analyses possible for different paradigms. Lastly, the
sub-event-horizon scale is precisely the place where the strong
resonance mechanism takes effect; and this greatly enhances the
observability of such signals for certain parameter space, even if the
vibrating field couples to the curvaton\footnote{In this context, we
  define the term ``curvaton" as the field that sources the leading
  scale-invariant power spectrum. It is one of the field directions in
  the low energy effective field space mentioned above. We have
  avoided using the term ``inflaton" because we are not just
  considering the inflationary paradigm, also because even for
  inflation the inflaton may not be the dominant source for the
  density perturbations.} only through gravity.

These signals show up as fine-structures in the density
perturbations. For the power spectrum, they appear as oscillatory
corrections to the leading scale-invariant shape. The
fraction $\Delta P_\zeta/P_{\zeta}$ is typically given
by~\cite{Chen:2011zf,Chen:2011tu}
\bea \frac{\Delta P_\zeta}{P_{\zeta}} = \Aw \left( \frac{2k}{\kr}
\right)^{-3+\frac{5}{2p}} \sin \left[ \frac{p^2}{1-p}
  \frac{2\ms}{\Hstar} \left( \frac{2k}{\kr} \right)^{1/p} + \varphi   \right].
\label{eq:fg_power}
\eea
The parameter we are interested in is $p$ -- the index of the
fingerprint of the primordial universe paradigm -- defined through the
scale factor evolution as
\bea a(t) = a(t_0) \left(\dfrac{t}{t_0}\right)^p.
\label{eq:powerlaw}
\eea

\begin{figure}
\begin{center}
\includegraphics[width=0.7\textwidth]{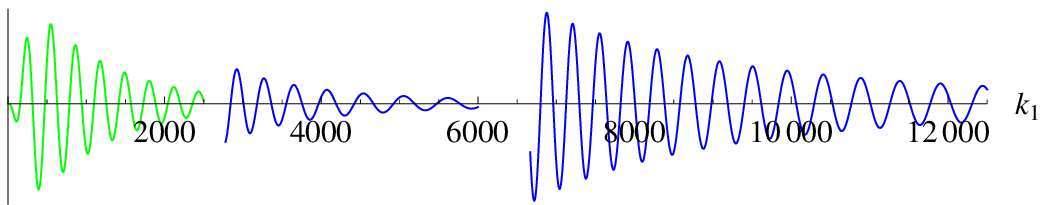}
\includegraphics[width=0.7\textwidth]{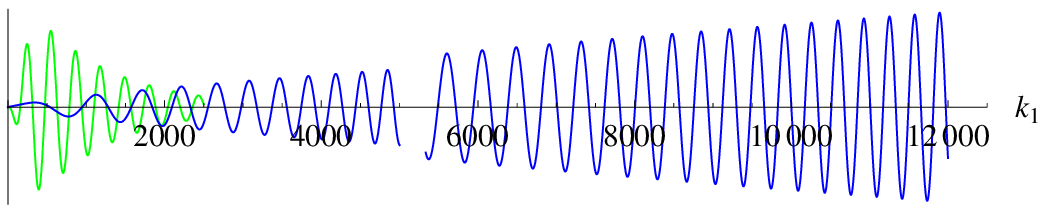}
\includegraphics[width=0.7\textwidth]{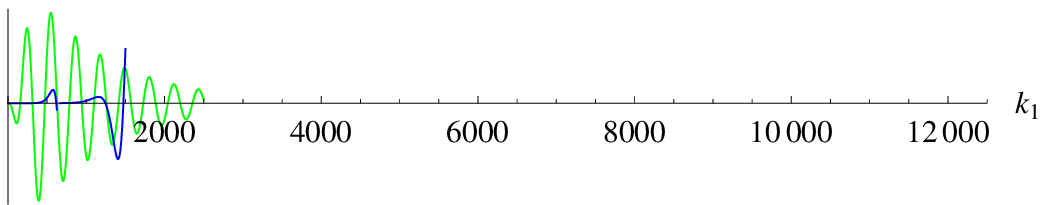}
\includegraphics[width=0.7\textwidth]{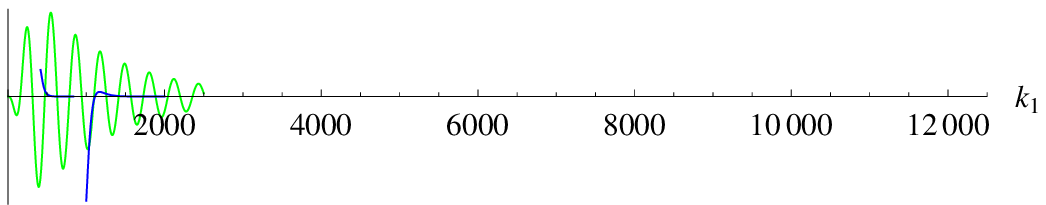}
\caption{``Fingerprints" of different paradigms induced by the
  ``standard clocks" in the power spectrum $\Delta
  P_\zeta/P_\zeta$. From top to bottom: $p=10$ (inflation), $p=2/3$
  (matter contraction), $p=0.2$ (slow contraction, or Ekpyrosis),
  $p=-0.2$ (slow expansion). The Green/light spectra are generated by
  a sharp feature at $k_0=100$ and have sinusoidal running; these
  features are not associated with ``clocks" and their behavior is
  similar for different paradigms. The blue/dark spectra correspond to
  two massive fields ($\ms= 25,60$) excited by this sharp feature and
  have resonant running; they contain the ``standard clocks" and can
  be used to measure $a(t)$ directly. The overall amplitudes of
  different signals have been normalized for clarity.}
\end{center}
\label{fig:fingerprints}
\end{figure}

Given $p$, whether the cosmological phase is expanding or contracting
is determined by the requirement that the quantum fluctuations should
exit the event-horizon during this epoch.  For example, $|p|>1$
corresponds to inflation, in which $p>1$ has slowly decreasing $H$
(with $t>0$) and $p<-1$ has slowly increasing $H$ (with $t<0$);
$p=2/3$ is the matter contraction phase; $0<p\ll 1$ is the ekpyrotic
(slowly contracting) phase; and $-1 \ll p <0$ describes a slowly
expanding phase. For recent reviews on these alternative scenarios,
see Refs.~\cite{Brandenberger:2012uj, Lehners:2011kr,
  Battefeld:2005av}. The parameter $\ms$ is the mass of the massive
field, $\kr$ is the first resonant mode excited by the oscillation,
and $\Hstar$ is the Hubble parameter at that moment. For the expanding
background, $p>1$ and $p<0$, lower $k$-modes resonate earlier and the
above formula applies to $2k>\kr$; for the contracting background,
$0<p<1$, larger $k$-modes resonate earlier and it applies to
$2k<\kr$. The corresponding patterns are illustrated in
Fig.~\ref{fig:fingerprints}.

The resonant running refers to the scale-dependence in the sine
function in Eq.~(\ref{eq:fg_power}). As we can see, the dependence of
this running pattern on the comoving momentum $k$ directly measures
the time dependence of the scale factor -- the two power-law functions
are inverse to each other.  This resonant running behavior is a very
robust signature for different paradigms because the ``zeros'' in
these oscillations cannot be changed by some potentially existing
curvaton-isocurvaton couplings during the multi-field evolution. These
zeros are the cosmological ticks imprinted by the standard clocks. For
examples, as we can see from Fig.~\ref{fig:fingerprints}, for an
expanding background such as inflation, the spacings between the ticks
increase in a specific way, while for a contracting background such as
the matter contraction, they decrease. In addition, the paradigms with
fast-evolving scale factors, such as the inflation and matter
contraction, have much more ticks than those with slowly-evolving
scale factors, such as ekpyrosis. The overall envelop behavior of
these oscillatory signals are less robust against the multi-field
complexities, but their overall scale-dependent trends are very clear.
Therefore by identifying the running patterns of these signals,
determined by the parameter $p$, we measure the fingerprint of a
specific primordial universe paradigm.

Specializing to inflation, there are many types of feature models that
can have interesting phenomenological consequences, and they give
different types of oscillatory signals in density perturbations. But
it is worth to emphasize that most of them cannot be used as the
distinguisher between the inflation and the alternative paradigms.  1)
The oscillations of massive fields we are interested in are induced by
certain sharp features. A sharp feature in itself generates a type of
oscillatory signals~\cite{Adams:2001vc, Chen:2006xjb,
  Achucarro:2010da, Adshead:2011bw, Hu:2011vr, Arroja:2011yu,
  Huang:2012mr, Arroja:2012ae, Avgoustidis:2012yc,Cespedes:2012hu,
  Abolhasani:2012px}. However, these signals are different from those
generated by the subsequent oscillations of the massive fields. Since
a sharp feature has only one ``click'', it does not provide a
``clock'' and the running pattern are universally varying
as\footnote{Here $k_0$ parameterizes the location of the sharp feature
  signal, and $\kr$ parameterizes the location of the resonant
  fingerprint signal. The two parameters are related, see
  Refs.~\cite{Chen:2011zf,Chen:2011tu}. For $p>0$, which include the
  inflation and fast-contraction paradigms, the locations of these two
  types of signals are hierarchically separated (see
  Fig.~\ref{fig:fingerprints}).} $\sim \sin(2k/k_0 + {\rm phase})$ for
all kinds of background evolution.  2) Some inflation models predict
small and repeated structures that can generate resonant feature
patterns~\cite{Chen:2008wn, Bean:2008na, Flauger:2009ab,
  Flauger:2010ja, Chen:2010bka, Leblond:2010yq, Behbahani:2011it,
  Aich:2011qv}. If these features are periodic, they generate the same
type of oscillatory behavior as from the standard clocks, but the
scale dependence of the envelop amplitudes are different. However,
such ``clocks'' are not standard. Instead, we are probing the
properties of these clocks by assuming the inflationary background.
Phenomenology of non-Bunch-Davies vacuum states in inflationary
background \cite{Niemeyer:2000eh, Martin:2000xs, Easther:2001fi,
  Easther:2001fz, Elgaroy:2003gq, Martin:2003sg, Kaloper:2003nv,
  Martin:2004iv, Easther:2004vq, Martin:2004yi, Schalm:2004qk,
  Hamann:2008yx} (or non-inflationary
background~\cite{Falciano:2008gt, Lilley:2011ag}) also belong to this
class -- the non-standard clock is now determined by the property of
the new physics scale.  3) The de Sitter inflationary phenomenology of
oscillating massive
fields~\cite{Burgess:2002ub,Shiu:2011qw,Gao:2012uq} can be easily
recovered by taking the large $p$ limit in (\ref{eq:fg_power}), in
which the power law dependence $p (2k)^{1/p}-p$ becomes the
logarithmic dependence $\ln (2k)$.

In this paper we study how the signals generated by these standard
clocks may be observed in the cosmic microwave background (CMB) data.
While we expect that such signals exist generically, their
observability depends on the parameter space of the models, the
state-of-art experimental technologies and data analyses methods. This
situation is similar to that for the tensor mode, but the signals are
of completely different types -- the features we are looking for are
fine-structures in the scalar density perturbations.  In addition, the
theory only predicts the patterns of the fingerprints, but not the
index, locations, frequencies, and amplitudes. Therefore, a search in
all the parameter space is needed. Using CMB data from the Wilkinson
Microwave Anisotropies Probe (WMAP) satellite, and Planck-like mock
data, we would like to find out which parts of the parameter spaces
are potentially observable and can be used to determine the range of
the fingerprint index $p$.

The paper is organized as follow. Using MCMC methods, we discuss in
Sec.~\ref{sec:wmap}, the constraints set by the WMAP satellite on the
eventual presence of signals given by
Eq.~(\ref{eq:fg_power}). Although we do not find evidence for
primordial fingerprints in the current data, there is a peculiar scale
around $10^{-2}\,\Mpc^{-1}$ at which such a power spectrum modulation
enhances the likelihood.

We then present forecasts for a typical Planck-like CMB experiment. As
the likelihood for superimposed oscillations in known to be
non-Gaussian~\cite{Martin:2004yi, Easther:2004vq}, we use MCMC methods
on generated mock data for various values of the resonance
parameters. This is the subject of Sec.~\ref{sec:planck}. We show that
there exists a ``high sensitivity zone'' for $10^{-2}\,\Mpc^{-1}
\lesssim k \lesssim 10^{-1}\,\Mpc^{-1}$ in which primordial
fingerprints can be detected\footnote{In terms of the multipole
  moments, using the approximate relation $\ell \simeq k \dang$ with
  $\dang \simeq 13.7\,\Gpc$~\cite{Page:2003fa}, the sensitivity zone
  corresponds to $200<\ell<1300$.}. Moreover, in presence of a low
signal-to-noise feature in the data, the first parameter to be
estimated would precisely be the wavenumber at which the modulation
occurs\footnote{In \cite{Chen:2011zf,Chen:2011tu} and here, the
  massive field is excited instantly by a sharp feature and decoupled
  from the curvaton afterward (i.e. coupled only
  gravitationally). This is the reason there are sharp edges near the
  starting wavenumber $\kf$ in (\ref{eq:fg_power}) and
  Fig.~\ref{fig:fingerprints}. This makes $\kf$ easier to be
  detected. If the (model-dependent) excitation and decoupling process
  happens more gradually, we expect some smoothing-out effect around
  the edges. The universal property of the fingerprints we emphasized
  here will show up as soon as the massive field and the curvaton can
  be treated as approximately decoupled.}, whereas frequency,
amplitude, phase and $p$ would remain unconstrained. The sensitivity
of Planck-like data allows a lower bound on the parameter
$|p|>1$, the sufficient condition for inflation, to be inferred for
amplitudes down to $ \max(\Delta P_\zeta/P_\zeta) \simeq 20\%$.
This threshold holds for low frequency signals whereas a full
reconstruction of the precise value of $p$ requires larger amplitudes,
typically greater than $50\%$. We also discuss how the
reconstruction is affected by the frequency and expansion parameter of
the underlying signal. We shall conclude in Sec.~\ref{sec:conclusion}.

\section{WMAP7 data}
\label{sec:wmap}

In this section, we consider the WMAP seven years
data~\cite{Komatsu:2010fb, Larson:2010gs, Jarosik:2010iu} for a flat
$\Lambda$CDM cosmological model with standard parameters, i.e. the
density of baryon $\OmegaB$, of cold dark matter $\OmegaDM$, the
optical depth $\tau$ and the Hubble parameter today $H_0$, or
equivalently, $\theta$ (which measures the angular size of the sound
horizon at last scattering~\cite{Lewis:1999bs}). Concerning the
primordial power spectrum, we consider only a scalar power-law power
spectrum having a ``fingerprints'' modulation as in
Eq.~(\ref{eq:fg_power}). Defining $\calP_\zeta = k^3/(2\pi^2) |
\zeta^2|$, we have
\begin{equation}
\label{eq:Pzeta}
  \calP_\zeta = \As
  \left(\dfrac{k}{k_*} \right)^{\ns-1} \left[1 + \Aw
    \left(\dfrac{k}{\kf}\right)^{-3 + 5/(2p)} \sin \left\{\omega
      \dfrac{p}{p-1} \left[p \left(\dfrac{k}{\kf} \right)^{1/p} - p
      \right] + \psi \right\} \right].
\end{equation}
The parameters $\As$ and $\ns$ are the usual amplitude and spectral
index while the pivot scale has been fixed to its fiducial value
$k_*=0.05\,\Mpc^{-1}$. We have moreover rescaled some of the
primordial parameters to reduce eventual degeneracies during the data
analysis. The scale at which the features is observed is $\kf \equiv
\kr/2$ and the frequency is now encoded in $\omega \equiv 2
\ms/\Hstar$. We have also rescaled the phase compared to
Eq.~(\ref{eq:fg_power}) as
\begin{equation}
\psi \equiv \varphi + 2\dfrac{\ms}{\Hstar} \dfrac{p^2}{p-1}\,,
\end{equation}
such that the large $p$ limit does not produce spurious correlations
between $\psi$ and $\omega$.

\subsection{Parameter space}

In presence of rapid oscillations, the derivation of the temperature
and polarization angular power spectra require some care. As discussed
in Ref.~\cite{Martin:2004iv}, the CMB transfer functions and the line
of sight integrals have to be estimated with a high precision to avoid
under-sampling of the oscillations. For this purpose, we have used a
modified version of the publicly available $\CAMB$
code\footnote{Available at:
  \url{http://theory.physics.unige.ch/~ringeval/upload/patches/features}}~\cite{Lewis:1999bs}. The
price to pay for accuracy is an unacceptable increase of the
computation time preventing any efficient exploration of the complete
parameter space. However, provided the modulation amplitude in the
$C_\ell$ remains small enough, and the frequency larger than the
typical acoustic oscillations, the parameter space associated with the
primordial parameter remains weakly correlated with the usual
cosmological parameters~\cite{Hamann:2008yx, Meerburg:2011gd}. For
this reason, and following Ref.~\cite{Martin:2004yi}, we freeze the
standard cosmological parameters to their best fit values, obtained
from a featureless power spectrum, and explore only the space of
primordial parameters. The technical advantage being that the CMB
transfer functions are computed only once. Concerning the WMAP data
analysis, we have used the publicly available MCMC code
$\COSMOMC$~\cite{Lewis:2002ah}, together with the likelihood provided
by the WMAP team~\cite{Jarosik:2010iu} and coupled to our modified
$\CAMB$ code. The sampling is thus performed over the primordial
parameters, i.e. $\ln(\As)$, $\ns$, $\Aw$, $\omega$, $p$, $\log(\kf)$
and $\psi$, starting with a flat prior distribution. The marginalized
posterior probability distributions are presented in the next section.

\subsection{Constraints on the fingerprint parameters}

As above-mentioned, the standard cosmological parameters have been
fixed to their fiducial values, i.e. $\OmegaB h^2=0.02286$, $\OmegaDM
h^2=0.115$, $\theta=1.044$ and $\tau=0.088$ ($h=0.71$). The MCMC
exploration has been stopped according to the $R-1$ statistics
implemented in $\COSMOMC$~\cite{Gelman:1992, Lewis:2002ah}, i.e. when
the estimated variance between different chains does not exceed
$1\%$. This number gives the typical error on the resulting posteriors
and have been reached for a number of samples around $300000$.

\begin{figure}
\begin{center}
\includegraphics[width=\textwidth]{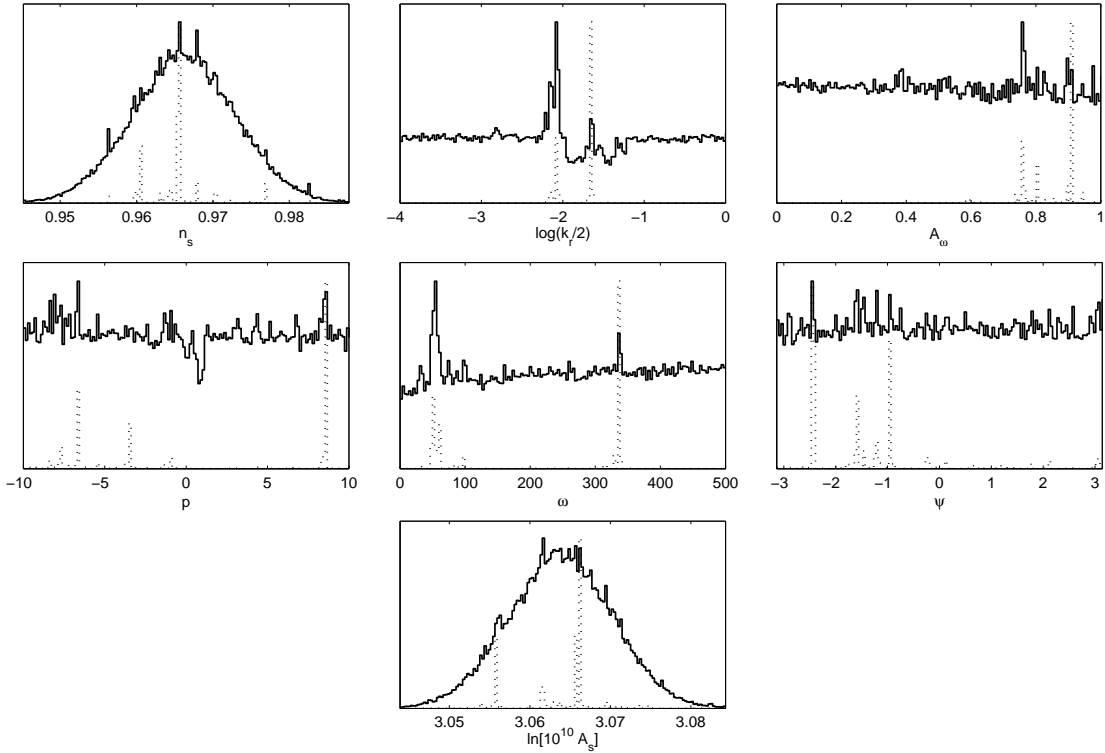}
\caption{Marginalized posterior probability distributions for the
  primordial parameters (solid), and mean likelihood (dotted),
  associated with the WMAP7 data (at fixed cosmological
  parameters). There is no bound on any fingerprint parameter,
  i.e. there is no evidence for, neither against, the presence of
  primordial features. Notice however the existence of two favoured
  scales around $\kf \simeq 10^{-2} \Mpc^{-1}$ (but with no
  statistical significance).}
\label{fig:fgwmap_1D}
\end{center}
\end{figure}

In Fig.~\ref{fig:fgwmap_1D}, we have plotted the marginalized
posterior probability distributions for the primordial parameters
(solid curve). Both $\As$ and $\ns$ are well constrained, as expected,
and centered at the best fit values found by considering only a
primordial power law spectrum~\cite{Komatsu:2010fb}. The variance is
somehow reduced, but this is the result of having fixed the
cosmological parameters. On the other hand, all the other parameters
exhibit flat distributions and therefore remain unconstrained. The
dotted curves in this figure show the mean
likelihood~\cite{Lewis:2002ah}, which typically traces the location of
the good fits. As discussed at length in Ref.~\cite{Martin:2004yi,
  Martin:2006rs}, the two distributions do not match when those better
fits require some amount of fine-tuning between the
parameters. Namely, they are located in small and correlated regions
of the parameter space such that their integrated weight on the
marginalized probability remains small. In Fig.~\ref{fig:fgwmap_1D},
this situation appears for multiple values of the parameters. For
instance, the distribution for $\ns$ exhibit various ``good fits''
whereas the overall probability remains of Gaussian shape. The weight
of each good fit can be assessed by the small deviation induced around
the Gaussian distribution. Along these lines, the distribution of
$\omega$ clearly exhibits two better fits for frequencies around $50$
and $300$, but the corresponding marginalized distribution does not
peak by more than one sigma in these locations. In fact, these two
frequencies are associated with the two favoured scales visible in the
distribution of $\kf$, which although not statistically significant,
are clearly distinguishable. As we will see in the following, the
scale $\kf$ is actually the most sensitive parameter to the actual
presence of a feature having a low signal-to-noise ratio.

The overall probability of having detected primordial fingerprints in
the WMAP7 data is given by the posterior of $\Aw$. Indeed, being
marginalized over all the other parameters, it gives the probability
density distribution of having a resonance of amplitude $\Aw$, for all
frequency, phase, location and power $p$. As one can check in
Fig.~\ref{fig:fgwmap_1D}, the $\Aw$ posterior is mostly flat meaning
that it is completely unconstrained. We conclude that there is no
evidence for primordial fingerprints in the WMAP data. On the other
hand, high values of $\Aw$ are not disfavoured such that there is also
no evidence against (up to the Occam's razor favouring the simplest
model~\cite{Trotta:2008qt}). This situation is in fact different than
unlocalized primordial oscillations, such as those coming from
non-standard vacuum initial conditions. In that case, large amplitudes
are disfavoured because the oscillations are spread over all
multipoles and, if too large, become incompatible with the observed
signal~\cite{Martin:2004yi, Martin:2006rs, Meerburg:2011gd}. Finally,
in Fig.~\ref{fig:fgwmap_1D}, one can notice that the posterior of $p$
is slightly reduced around small positive values. These models
correspond to fast contraction, which, compared to the other expansion
paradigms, have tendency to produce more widely spread oscillations
in the $C_\ell$. As a result, too large amplitudes are not admissible
and this region ends up being slightly disfavoured after
marginalization. These models will be discussed in more details in
Sec.~\ref{sec:beyond}.

In view of these results, it is instructive to discuss how much future
CMB data can constrain the typical signal associated with features. In
particular, do the peaks in the $\kf$ distribution could be
interpreted as hints of primordial fingerprints? In the next section,
we present forecasts for an ideal Planck-like CMB experiment, using
similar MCMC methods on mock data. We will see that $\kf$ is indeed
the most sensitive parameter to an underlying localized
modulation\footnote{Let us notice however that such a modulation, in a
  realistic experiment, may also be generated by some residual
  colored noise.}. We will find out the minimum amplitude
$\Aw$ detectable for different fingerprints as we vary the frequency
$\omega$ and fingerprint index $p$.

\section{Planck-like CMB data}

\label{sec:planck}
\begin{figure}
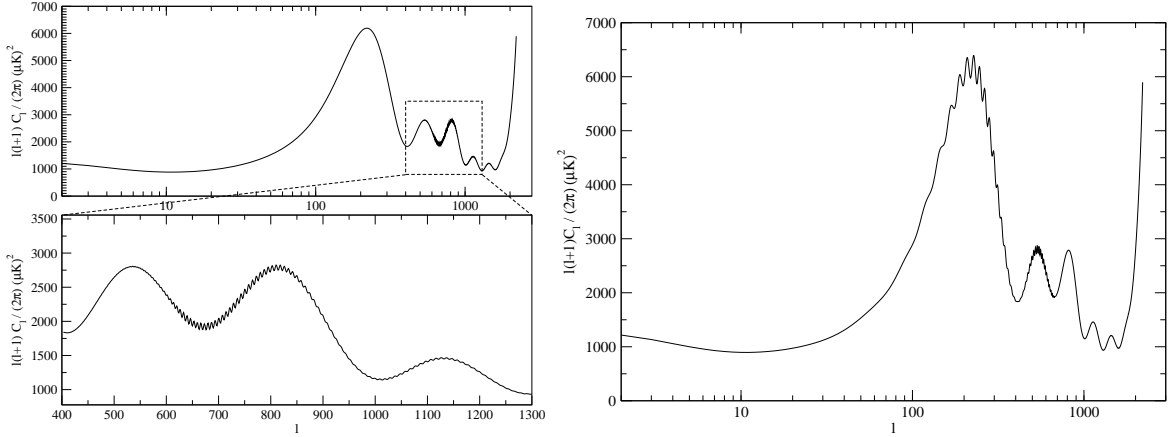

\begin{center}
\includegraphics[width=0.46\textwidth]{figs/cls_aw050_w600}
\includegraphics[width=0.52\textwidth]{figs/cls_aw040_w200_p2o3}
\caption{Temperature angular power spectrum, plus noise, for an
  inflationary feature (left panel) located at
  $\fiducial{\kf}=0.05\,\Mpc^{-1}$, $\fiducial{\Aw}=0.5$,
  $\fiducial{\omega}=600$, $\fiducial{p}=8$ (inflation) and
  $\fiducial{\psi}=0$. The right panel shows a feature at the same
  location but during a fast contracting era having
  $\fiducial{\kf}=0.05\,\Mpc^{-1}$, $\fiducial{\Aw}=0.4$,
  $\fiducial{\omega}=200$, $\fiducial{p}=2/3$ (matter contraction) and $\fiducial{\psi}=0$
  (see Sec.~\ref{sec:beyond}).}
\label{fig:cls_aw050}
\end{center}
\end{figure}

Given a fiducial set of cosmological parameters $\{\fidlambda_a \}$,
the easier method to forecast their expected errors would be to use
the Fisher matrix formalism which merely consists as approximating the
likelihood by a Gaussian around the best fit
location~\cite{Tegmark:1996bz}. However, as discussed in
Refs.~\cite{Martin:2004yi, Martin:2006rs, Easther:2004vq} the
likelihood associated with superimposed oscillations in the $C_\ell$
is non-Gaussian and can be multi-valued such that Fisher matrix method
can only be applied for a high signal-to-noise detection. For this
reason, we prefer in the following a MCMC exploration based on mock
data containing various fingerprint oscillatory
patterns~\cite{Easther:2004vq}.

\subsection{Mock power spectrum and likelihood}

Forecasts can be made through an MCMC exploration of the parameter
space provided one specifies the likelihood. The mock data
$\{\fidCb_\ell\}$ are assumed to be associated with a set of fiducial
cosmological and primordial parameters $\{\fidlambda_a\}$ and one
needs to specify the likelihood of the theoretical
$\{\Cb_\ell(\lambda_a)\}$ tested. For a full sky analysis, assuming
isotropic white noise for each pixel and Gaussian statistics, one can
show that the sampling distribution followed by the $C_\ell$ is a
Wishart distribution~\cite{Bond:1998qg, Perotto:2006rj,
  Percival:2006ss, Hamimeche:2008ai}. Including polarization yields,
up to a normalization constant~\cite{Colombo:2008ta}
\begin{equation}
\label{eq:like}
-2 \ln \like{\Db_\ell}{\fidDb_\ell} = \sum_\ell (2\ell+1)
\left[ \trace\left(\fidDb_\ell\cdot \Db_\ell^{-1}\right) - \ln\left|
    \fidDb_\ell\cdot \Db_\ell^{-1}\right| - 3 \right].
\end{equation}
The matrix $\Db_\ell = \{D_\ell^{XY}\}$ where $X$, $Y$ stand for
temperature and polarization variables, $T$, $E$ and $B$. The spectra
$D_\ell^{XY}$ include a white noise component compared to the angular
power spectra $C_\ell^{XY}$ and are defined by
\begin{equation}
D^{XY}_\ell \equiv C^{XY}_\ell + \dfrac{\Cnoise^{X}}{(B^{X}_\ell)^2}\,,
\end{equation}
where $B^{X}_\ell$ is the beam response. For a Planck-like experiment,
we have chosen a Gaussian beam with a full width at half maximum
(fwhm) of $7'$. The noise power for temperature has been set to
$\Cnoise^{T}=2 \times 10^{-4}\,\muK^2$ and for polarization to
$\Cnoise^{E} \simeq \Cnoise^{B} \simeq 2
\Cnoise^{T}$~\cite{LAMARRE-2010-535714}. From our modified version of
the $\CAMB$ code, and the above noise specification, the $D^{XY}_\ell$
can be computed for any input value of the cosmological and primordial
parameters $\{\lambda_a\}$. Two examples of the temperature angular
power spectrum are represented in Fig.~\ref{fig:cls_aw050}. Note that
the oscillation amplitudes are greatly reduced in the temperature
angular power spectrum comparing to those in the primordial power
spectrum.

Using MCMC sampling with the likelihood of Eq.~(\ref{eq:like}) allows
to extract the posterior probability distribution for each
``measured'' parameter $\lambda_a$ given the fiducial model
$\{\fidlambda_a\}$. For the same reasons discussed in
Sec.~\ref{sec:wmap}, we have fixed the cosmological parameters to
their best fit value, obtained from a standard power-law primordial
power spectrum, and ran the MCMC exploration only in the primordial
parameter space. The standard cosmological fiducial parameters are the
same as in Sec.~\ref{sec:wmap}, plus $\fiducial{\ns}=0.97$ and
$\ln(10^{10}\fiducial{\As})=3.166$. In the following, we consider
various fiducial values for the fingerprint parameters and discuss how
well they can be reconstructed.

\subsection{No feature: the sensitivity zone}

\begin{figure}
\begin{center}
\includegraphics[width=\textwidth]{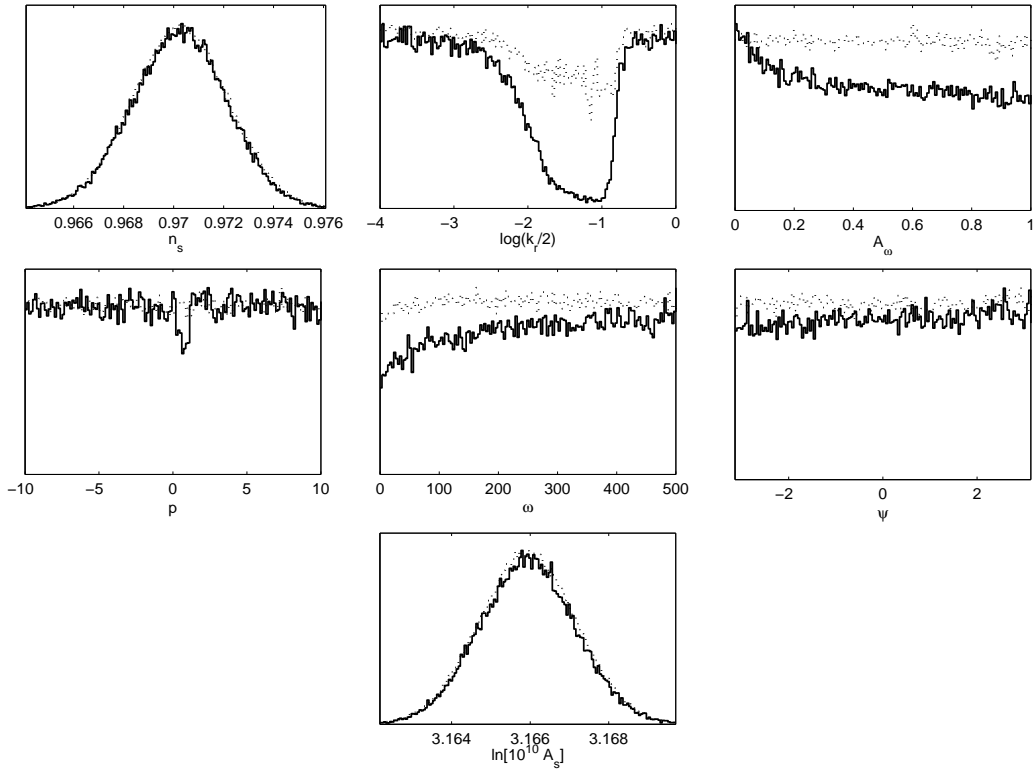}
\caption{Marginalized posterior probability distributions for the
  primordial parameters (solid), and mean likelihood (dotted),
  associated with the mock Planck-like data when there is not any
  feature present. Notice the posterior of $\kf$ which inversely
  traces the feature sensitivity region.}
\label{fig:aw000_1D}
\end{center}
\end{figure}

The first fiducial model considered has no feature,
i.e. $\fiducial{\Aw}=0$. Practically, the MCMC chains are run as
specified in Sec.~\ref{sec:wmap}, with the same convergence criteria,
i.e. the chains are stopped when the expected error on the
marginalized distributions does not exceed a few percents. In
Fig.~\ref{fig:aw000_1D}, we have represented the marginalized
posterior distributions obtained from the MCMC exploration of the
primordial parameter space.

\begin{figure}
\begin{center}
\includegraphics[width=0.8\textwidth]{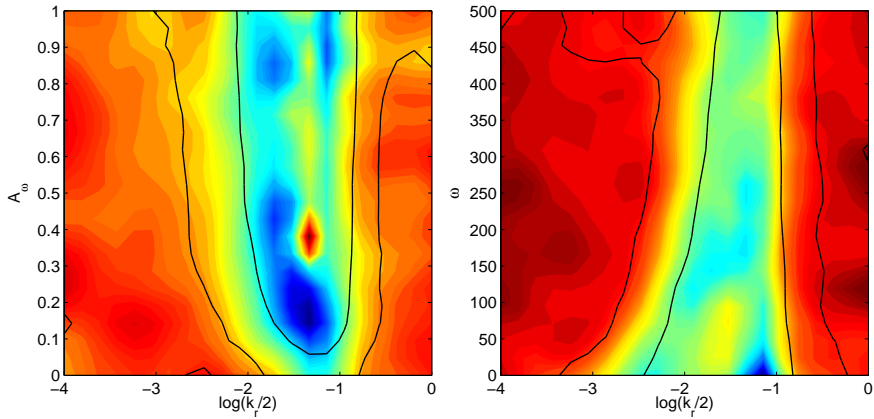}
\caption{One- and two-sigma confidence intervals in the planes
  $(\log\kf,\Aw)$ and $(\log\kf,\omega)$ and mean likelihood (shading)
  for vanishing fingerprints in the mock data. These domains trace the
  sensitivity region of Planck-like CMB data. For $\kf \gtrsim
  10^{-1}\,\Mpc^{-1}$ or $\kf \lesssim 10^{-2}\,\Mpc^{-1}$, localized
  resonances are hardly detectable.}
\label{fig:aw000_2D}
\end{center}
\end{figure}

The distribution for the standard primordial parameters, $\ns$ and
$\As$, are reconstructed, as expected for well constrained
parameters. Almost all fingerprint parameters exhibit a flat
distribution and are unconstrained. The slight deviations
between mean likelihood and marginalized posteriors for $\Aw$,
$\omega$ and $\kf$ are the signatures of correlations. In
Fig.~\ref{fig:aw000_2D}, we have plotted the two-dimensional one- and
two-sigma confidence intervals as well as the two-dimensional mean
likelihood (shading). This plot shows that there is a strongly
disfavoured region for $\kf$ between $10^{-2}\,\Mpc^{-1}$ and
$10^{-1}\,\Mpc^{-1}$. In this domain, Planck-like data are sensitive
to the presence of features and as our fiducial model has no feature,
this region is disfavoured. Conversely, the other domains are poorly
constrained. On smaller scales the noise starts to dominate whereas on
larger scales any resonance pattern is smoothed out by the CMB
transfer functions~\cite{Martin:2004yi, Martin:2006rs}. The
one-dimensional distributions of Fig.~\ref{fig:aw000_1D} end up being
flat because, for any $\Aw$ (or $\omega$) value, there exists a scale
$\kf$ for which the model can be made undetectable within a
Planck-like accuracy CMB experiment.

\subsection{Fingerprints of inflation}

\label{sec:inf}

As a starting point, we consider an inflation model with
$\fiducial{p}=8$, having a strong feature\footnote{Despite of the
  large amplitude in the primordial power spectrum, in slow-roll
  inflation this example corresponds to a transfer of $\sim 3\%$ of the
  inflaton kinetic energy to the massive field
  \cite{Chen:2011zf}.} of amplitude $\fiducial{\Aw}=0.5$, at a
high frequency $\fiducial{\omega}=600$, and located in the middle of
the detectable zone, i.e. $\fiducial{\kf} = 0.05\,\Mpc^{-1}$
($\psi=0$). Such a model has an angular temperature power spectrum
represented in Fig.~\ref{fig:cls_aw050} (the polarization spectra are
not represented but exhibit a similar behaviour).

\begin{figure}
\begin{center}
\includegraphics[width=\textwidth]{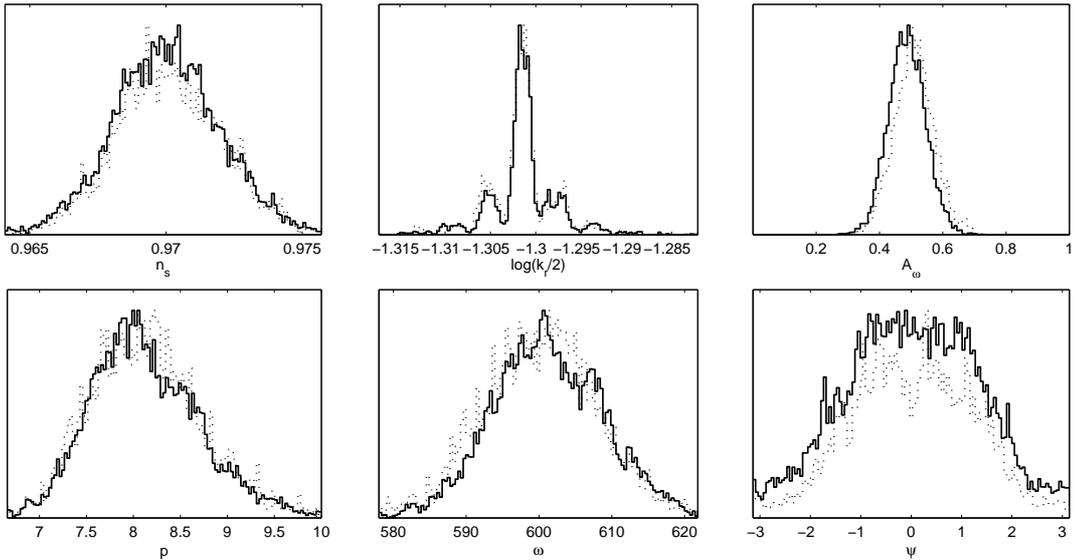}
\caption{Marginalized posterior probability distributions for the
  primordial parameters (solid), and mean likelihood (dotted), for
  typical Planck-like mock CMB data having a feature with
  $\fiducial{\Aw}=0.5$, $\fiducial{\omega}=600$, $\fiducial{\psi}=0$,
  $\fiducial{p}=8$ and $\fiducial{\kf} = 0.05\,\Mpc^{-1}$ (same as in
  Fig.~\ref{fig:cls_aw050}). Being in the high sensitivity zone, all
  fingerprint parameters are well reconstructed, especially the scale
  $\kf=\kr/2$. For such a strong signal, the fingerprint parameter $p$
  can even be precisely measured, in addition to the conclusion that
  $|p|>1$.}
\label{fig:aw050_w600}
\end{center}
\end{figure}

Running a MCMC exploration on the primordial parameters against such a
model yields the marginalized distributions of
Fig.~\ref{fig:aw050_w600}. The power spectrum amplitude $\As$
is always well-constrained and its posterior remains identical to the
one of Fig.~\ref{fig:aw000_1D}. It is not represented in the
following. The phase $\psi$ is poorly recovered whereas all the other
fingerprint parameters are well determined. Their standard deviation
are $\Delta \Aw = 0.08$, $\Delta \omega=43$, $\Delta \Psi = 1.2$ and
$\Delta p =2.9$ showing that such a feature would indeed allow to
probe the expansion rate through $p$. The most sensitive parameter
remains however the wavenumber scale as we find its standard deviation
to be $\Delta \kf/\kf =4 \times 10^{-3}$.

\begin{figure}
\begin{center}
\includegraphics[width=\textwidth]{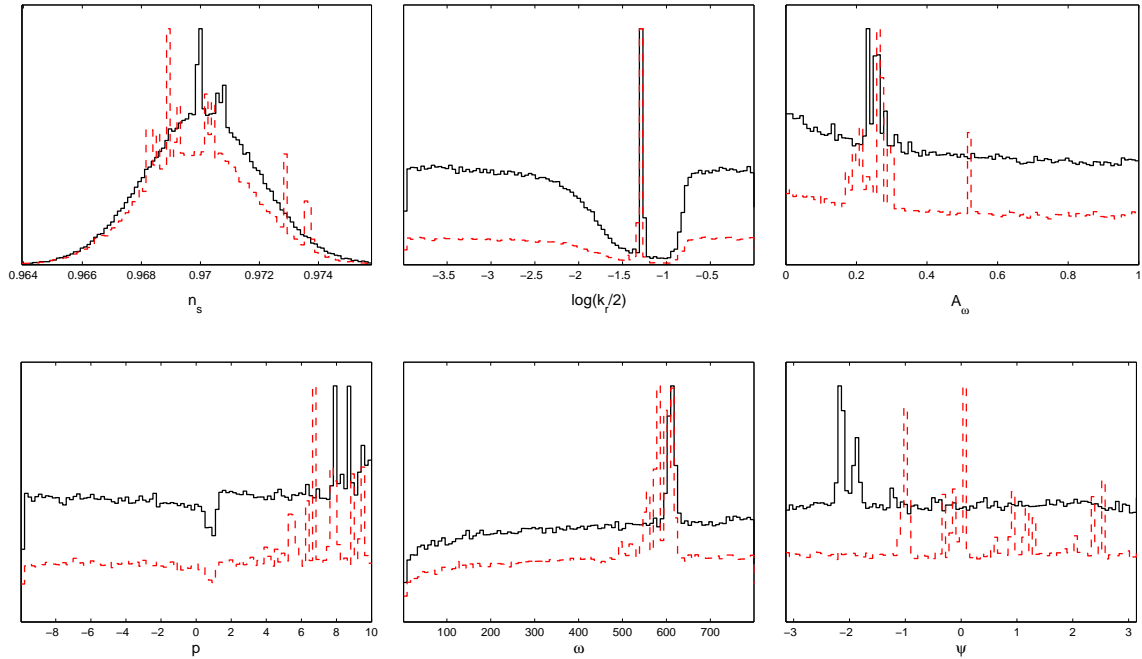}
\caption{Marginalized posteriors obtained by reducing the feature
  amplitude to $\fiducial{\Aw}=0.4$ (red dashed) and
  $\fiducial{\Aw}=0.3$ (black solid). The other parameters are the
  same as in Fig.~\ref{fig:aw050_w600}. Already for
  $\fiducial{\Aw}=0.4$, the posterior for $\Aw$ gets noisy showing
  that such a feature is no longer detectable. However, frequency and
  scale are still felt and are the most sensitive
  parameters. Amplitude, phase and the expansion parameter $p$
  requires a stronger signal to be inferred (see
  Fig.~\ref{fig:aw050_w600}).}
\label{fig:awXXX_w600}
\end{center}
\end{figure}

In order to test the sensitivity of Planck-like CMB data with respect
to the amplitude, we have redone the same analysis for various
fiducial amplitudes $\fiducial{\Aw}$, all the other fiducial
primordial parameters being unchanged. The marginalized posteriors are
represented in Fig.~\ref{fig:awXXX_w600} and shows that for
$\fiducial{\Aw} \lesssim 0.4$, any detection becomes unlikely and
impossible for $\fiducial{\Aw}<0.3$. Let us notice that already for
$\fiducial{\Aw}=0.4$, amplitude, phase and the expansion parameter $p$
are poorly, if not recovered. This suggests that, for a high frequency
fingerprint, the presence of a strong feature, as the one discussed
previously, is crucial for probing the $p$ parameter. There exists a
very sharp line in terms of the value of $\fiducial{\Aw}$ between the
signals that can be reconstructed and those cannot. On the other hand,
the parameters such as the frequency $\omega$ and the feature scale
$\kf$ still let some imprints, down to $\fiducial{\Aw}=0.3$.

\begin{figure}
\begin{center}
\includegraphics[width=0.8\textwidth]{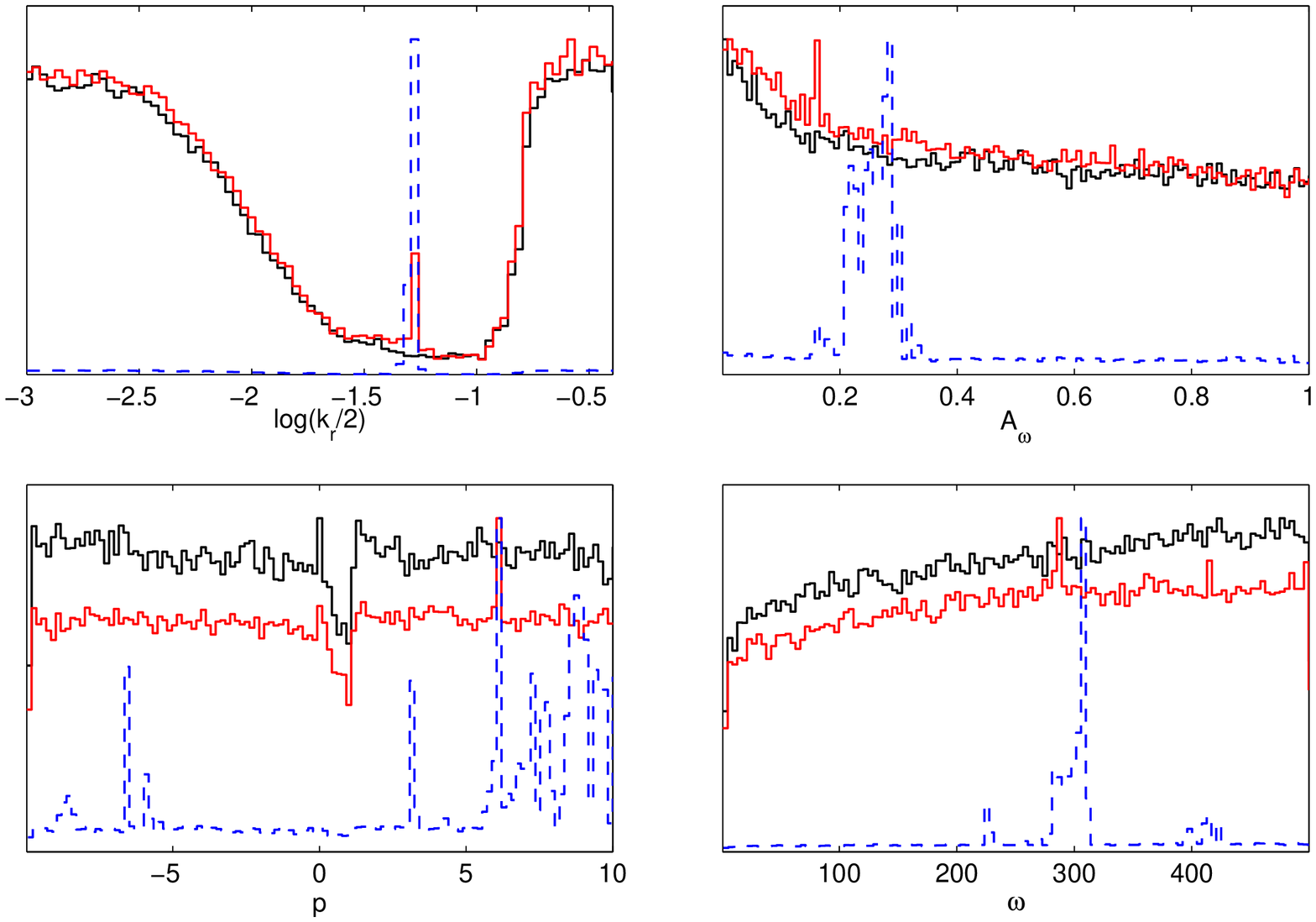}
\caption{Same as Fig.~\ref{fig:awXXX_w600} but for a reduced frequency
  $\fiducial{\omega}=300$. The sensitivity is slightly increased as
  signals down to $\fiducial{\Aw} \simeq 0.3$ are now detectable in
  the posterior for $\Aw$. Only the constrained fingerprint parameters
  are represented for three fiducial amplitudes $\fiducial{\Aw}=0.3$
  (dashed blue), $\fiducial{\Aw}=0.2$ (solid red) and
  $\fiducial{\Aw}=0.1$ (thick black).}
\label{fig:awXXX_w300}
\end{center}
\end{figure}

The effect from changing the fiducial frequency in the mock data is
typical of any other resonant pattern: the CMB transfer function
smoothing out rapid oscillations, at constant $\fiducial{\Aw}$, higher
frequencies produce a lower signal in the CMB~\cite{Martin:2004iv}. In
Fig.~\ref{fig:awXXX_w300}, we have plotted the posteriors obtained for
a fiducial model having $\fiducial{\omega}=300$ and for various values
of the amplitude $\fiducial{\Aw}$. The behaviour is exactly the same
as for $\fiducial{\omega}=600$, but the sensitivity to
$\fiducial{\Aw}$ is increased because the resonances are less smoothed
out by the CMB transfer function. So signals with lower amplitudes
become slightly more accessible.

\begin{figure}
\begin{center}
\includegraphics[width=\textwidth]{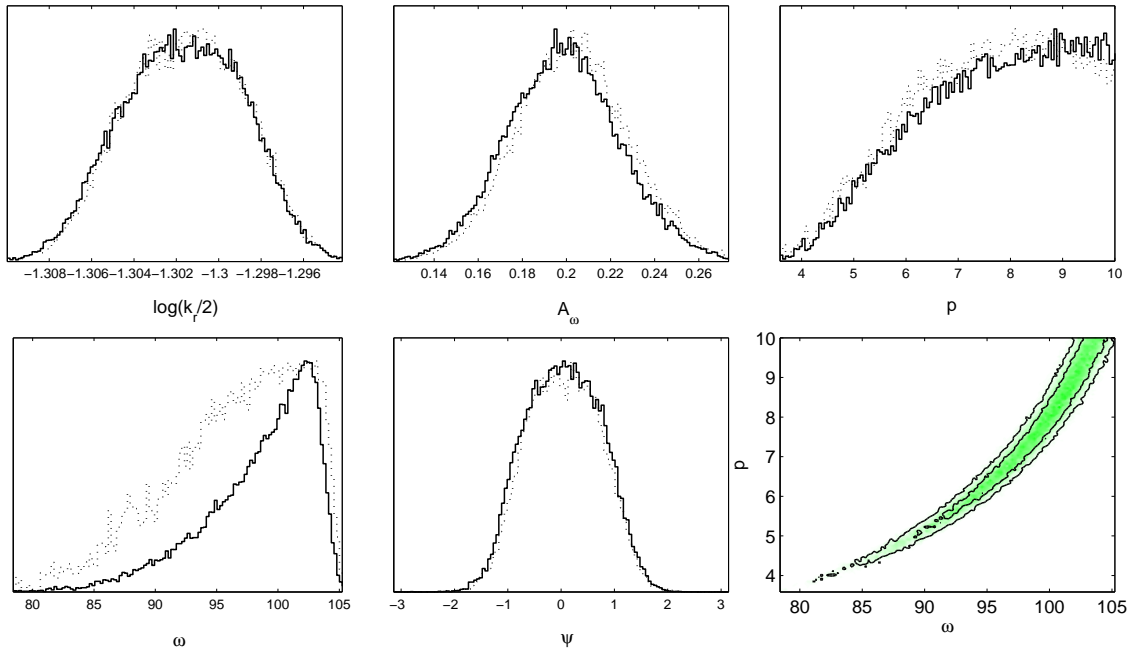}
\caption{Posteriors obtained with a small inflationary feature at low
  frequency: $\fiducial{\Aw} \simeq 0.2$ and
  $\fiducial{\omega}=100$. Although low frequencies render smaller
  oscillation amplitude detectable, the $p$ value cannot be
  reconstructed but still indicates an inflationary paradigm. At lower
  frequencies, for instance $\fiducial{\Aw}=50$, the posterior of $p$
  extends to negative values such that distinguishing the inflationary
  paradigm from another expansion era becomes more difficult (see
  text). The lower right panel shows the one- and two-sigma confidence
  intervals of two-dimensional posterior in the plane $(\omega,p)$:
  the two parameters are correlated.}
\label{fig:aw020_w100}
\end{center}
\end{figure}

We have further tested examples with relatively low frequencies.  In
Fig.~\ref{fig:aw020_w100} we have shown the posteriors for an
underlying fingerprint with $\fiducial{w}=100$ and
$\fiducial{\Aw}=0.2$. We can see that, for low frequency, a
fingerprint would be detected by Planck with much lower amplitude. For
example, the posterior for $\Aw$ is sharply peaked at the fiducial
value with a standard deviation of $\Delta \Aw=0.03$. While most of
the parameters can be reconstructed as before, there is an interesting
exception of the posterior of $p$. This parameter can now only be
constrained from below: $p>5$ at $95\%$ of confidence. The loss of
accuracy on $p$ also affects the determination of $\omega$ as both
parameters are degenerated (see the two-dimensional posterior in
Fig.~\ref{fig:aw020_w100}). This is not difficult to explain. In the
large $|p|$ limit, the fingerprint profile approaches to a unique de
Sitter limit, so fingerprints with large $p$ values tend to be
degenerate. With lower frequency, this degeneracy becomes more
effective since the ``standard clock'' is running slower and there are
fewer ``ticks'' available to reconstruct the exact parameter $p$.
However for our purpose, the exact value of $p$ is not the most
important one. As long as we can demonstrate $|p| \gg 1$, we would be
able to identify the inflation as the underlying paradigm.
Interestingly, Fig.~\ref{fig:aw020_w100} shows that the posterior for
$p$ indeed unambiguously indicates that this is an inflationary
paradigm, although recovering $\fiducial{p}=8$ is no longer
possible. We have also tested a lower frequency signal having
$\fiducial{\omega}=50$ (and $\fiducial{\Aw}=0.2$, figures not
represented). The posteriors are very similar to those of
Fig.~\ref{fig:aw020_w100}, the amplitude and frequency are peaked at
their fiducial value. The exact $p$ value cannot be reconstructed as
well, and now the degeneracies between $(\omega,p)$ are extended to
negative $p$ values (including the other inflationary branch
$p<-1$). The bound for inflation, $|p|>1$, can still be established
but by not more than two-sigma as the $99\%$ confidence region
includes other paradigms. Further lowering the frequency however will
not help to reduce the minimum detectable amplitude. This is because,
with fewer oscillations, the fingerprints start to be confused with
the acoustic oscillations. They sometimes do not even appear as
oscillations, but only as deformations of various acoustic peaks.

\begin{figure}
\begin{center}
\includegraphics[width=\textwidth]{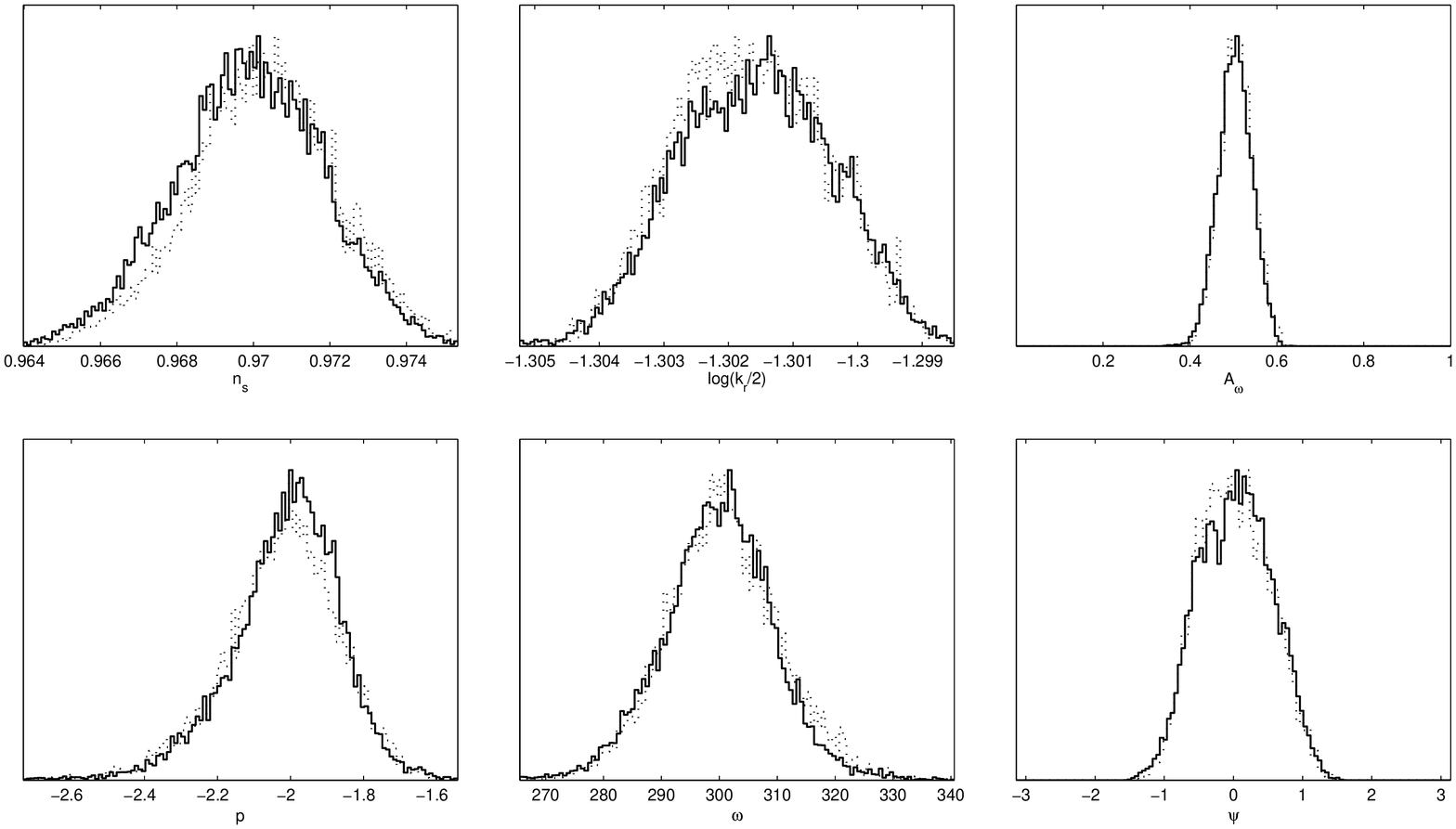}
\caption{Primordial posteriors for an expanding model having
  $\fiducial{p}=-2$, with $\fiducial{\Aw}=0.5$ and
  $\fiducial{\omega}=300$. The expansion parameter $p$ is recovered at
  a standard deviation of $\Delta p=0.15$.}
\label{fig:aw050_w300_pm2}
\end{center}
\end{figure}

In order to explore different possibilities, we have also tested a
rather unusual inflation model with $\fiducial{p}=-2$ and
$\fiducial{\omega}=300$, all the other fiducial parameters being as
before. This is still inflation in the sense that $|p|>1$, but unusual
because the expansion rate is much slower than the de Sitter space and
the Hubble parameter is increasing. As Eq.~(\ref{eq:fg_power}) shows,
the $k$-dependency is such that the ``instantaneous'' frequency
decreases faster than a logarithm for $k/\kf \gg 1$. From the CMB
point of view, it means that the signal is less damped by the transfer
functions and the oscillatory pattern is spread over larger multipoles
than for the inflationary paradigm ($p\gg 1$). In
Fig.~\ref{fig:aw050_w300_pm2}, we have represented the marginalized
posteriors obtained by a MCMC analysis. Compared to the case
$\fiducial{p}=8$, fingerprint parameters are more constrained due to
the larger CMB signal, this is particularly clear for the phase
$\psi$. Varying amplitude and frequency reproduces the same
qualitative behaviour discussed before, namely one would find that all
features disappear for $\fiducial{\Aw}<0.1$. Again, $\kf$ remains the
most sensitive parameter as we find his posterior well peaked down to
$\fiducial{\Aw}=0.2$.

\subsection{Beyond the inflationary paradigm}
\label{sec:beyond}

In this section, we discuss the detectability of the fingerprints
within alternatives to inflation. We consider three categories:
slow-contraction models ($0<p\ll 1$), slow-expansion models ($-1 \ll p
<0$) and fast-contraction models ($p \lesssim 1$). Fast-expanding but
non-inflationary cases ($-1<p<0$ and $p\sim -1$) are similar to
the $\fiducial{p}=-2$ case considered in the previous section.

\begin{figure}
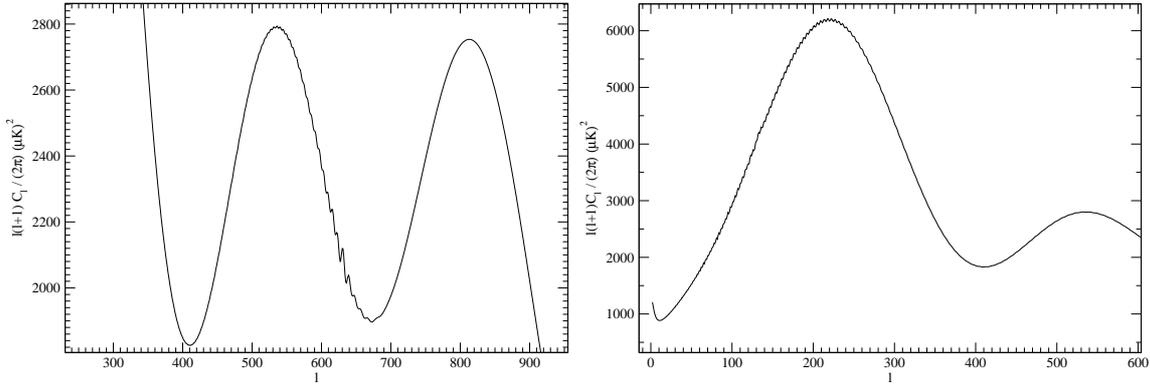

\begin{center}
\hspace{-0.9cm} \includegraphics[width=0.48\textwidth]{figs/cls_aw050_w6000_p01}
\includegraphics[width=0.485\textwidth]{figs/cls_aw080_w300_k001}
\caption{Two types of undetectable features for a Planck-like
  experiment. Left panel: zoom on the second and third acoustic peaks
  for a slow contracting model $\fiducial{p}=0.1$,
  $\fiducial{\Aw}=0.5$ and $\fiducial{\omega}=6000$. The resonances
  are before the feature scale $\kf=0.05\,\Mpc^{-1}$ but remain very
  localized due to the small value of $p$. Right panel: an inflation
  model, $\fiducial{p}=8$, having a strong feature
  $\fiducial{\Aw}=0.8$ washed out under cosmic variance by the large
  scale CMB transfer functions. It has $\kf=0.01\,\Mpc^{-1}$ and
  $\fiducial{\omega}=300$ (see Sec.~\ref{sec:nozone}).}
\label{fig:cls_p01}
\end{center}
\end{figure}

To test the slowly contracting models, we consider the case
$\fiducial{p}=0.1$. The effective resonance frequency being given by
$\omega p^2$, superimposed oscillatory patterns end up being of
observable frequency for $\omega \gg 1/p^2$. For this reason, we have
considered a fiducial model having $\fiducial{\omega}=6000$, its
temperature angular power spectrum is represented in
Fig.~\ref{fig:cls_p01} for $\fiducial{\Aw}=0.5$. The small value of
$p$ makes the oscillation pattern very localized around $\kf$ and this
model ends up being unobservable. Even for an unrealistic maximal
modulation amplitude ($\fiducial{\Aw}=1$), the MCMC analysis does not
allow the reconstruction of any fingerprint parameters.

The case of slowly expanding models is very close, their fingerprints
exhibit similarity to the slowly contracting models which would affect
their detectability (see the last two figures in
Fig.~\ref{fig:fingerprints}).

\begin{figure}
\begin{center}
\includegraphics[width=\textwidth]{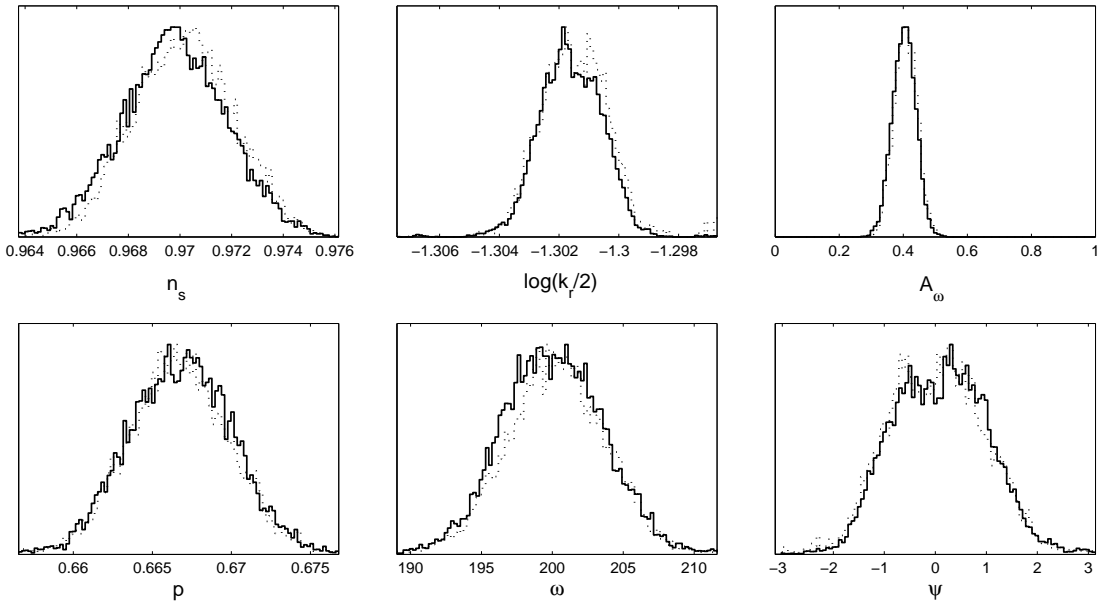}
\caption{Expected marginalized posteriors for a feature generated
  during a fast contracting era and having $\fiducial{\Aw}=0.4$,
  $\fiducial{\omega}=200$, $\fiducial{p}=2/3$ (the temperature power
  spectrum is plotted in Fig.~\ref{fig:cls_aw050}).}
\label{fig:p2o3}
\end{center}
\end{figure}

Faster contracting models, having larger values of $\fiducial{p}$,
should not suffer from this problem, as for instance a matter
contraction with $\fiducial{p}=2/3$. On the contrary, they exhibit
widespread oscillations in the primordial power spectra for $k<\kf$
but, as seen in Eq.~(\ref{eq:fg_power}), their ``instantaneous''
frequency increases dramatically. Depending on the values of $\kf$, if
the frequency $\omega$ becomes too large, the CMB transfer functions
acting as a low-pass filter strongly damp the oscillatory tail at
large multipoles. As a result, the fingerprints may no longer be
visible above some $\ell$ value, which could be lower than $\ellf$,
thereby rendering parameter reconstruction difficult. These models can
therefore be ``visible'' only for not too large frequencies, but this
also implies the existence of a few oscillations in the $C_\ell$
having a stronger amplitude than for inflation. This is illustrated in
the right panel of Fig.~\ref{fig:cls_aw050} where we have plotted the
angular power spectrum associated with the feature
$\fiducial{\omega}=200$ and $\fiducial{\Aw}=0.4$. The resonance
patterns show up at the top of the first peak whereas the feature
scale is located at much smaller scales $\ellf\simeq 700$. In fact,
such a characteristic explains why those models are a bit disfavoured
by the WMAP7 data: compared to other early universe paradigms, at same
amplitude and frequency, those $p$ values are associated with slightly
too large oscillations in the $C_\ell$. Concerning the Planck
forecasts, we have represented in Fig.~\ref{fig:p2o3} the posteriors
obtained from the MCMC analysis. For the same fiducial amplitude
$\fiducial{\Aw}=0.4$, all parameters are well reconstructed and in
particular the expansion index $p$. The above-mentioned sensitivity
to the model parameters renders smaller amplitudes hardly detectable
for Planck-like data. We have tested a smaller fiducial amplitude of
$\fiducial{\Aw}=0.3$ (figures not represented) for which the posterior of
$\Aw$ appears very noisy and not clearly peaked. However, and contrary
to the inflation paradigms, the posterior of $p$ still indicates a
contracting model because, even noisy, the oscillatory patterns remain
typical of a fast contracting background. In that situation, $p$ ends
up being a more sensitive parameter than the amplitude. For
$\fiducial{\Aw}\lesssim 0.2$, all hints for a signal are lost and the
posteriors are identical to those of Fig.~\ref{fig:aw000_1D}.

\subsection{Outside the sensitivity zone}

\label{sec:nozone}

As one may expect, if the primordial feature occurs at a scale $\kf$
which is well outside the sensitivity region, it cannot produce a
large enough signal in the CMB to be clearly distinguished. To check
how the transition occurs, we have performed a MCMC exploration
for the same fiducial model as in Sec.~\ref{sec:inf}
($\fiducial{p}=8$, $\fiducial{\omega}=300$) but with a scale
$\kf=0.01\,\Mpc^{-1}$. Even for an amplitude $\fiducial{\Aw}=0.8$, the
marginalized posteriors for all primordial parameters are identical to
those of Fig.~\ref{fig:aw000_1D}, i.e. as if no feature were
present. As explained before, this is the result of the strong
smoothing by the CMB transfer functions, which are very efficient on
the largest scales. Such a smoothing is visible on the fiducial power
spectra as the resonance patterns remain under the cosmic variance
(see right panel of Fig.~\ref{fig:cls_p01}). Let us mention that not
increasing the numerical accuracy for the transfer functions,
i.e. using $\CAMB$ at its default numerical precision, may produce
spurious enhanced signals in the CMB~\cite{Martin:2004iv}.

\begin{figure}
\begin{center}
\includegraphics[width=\textwidth]{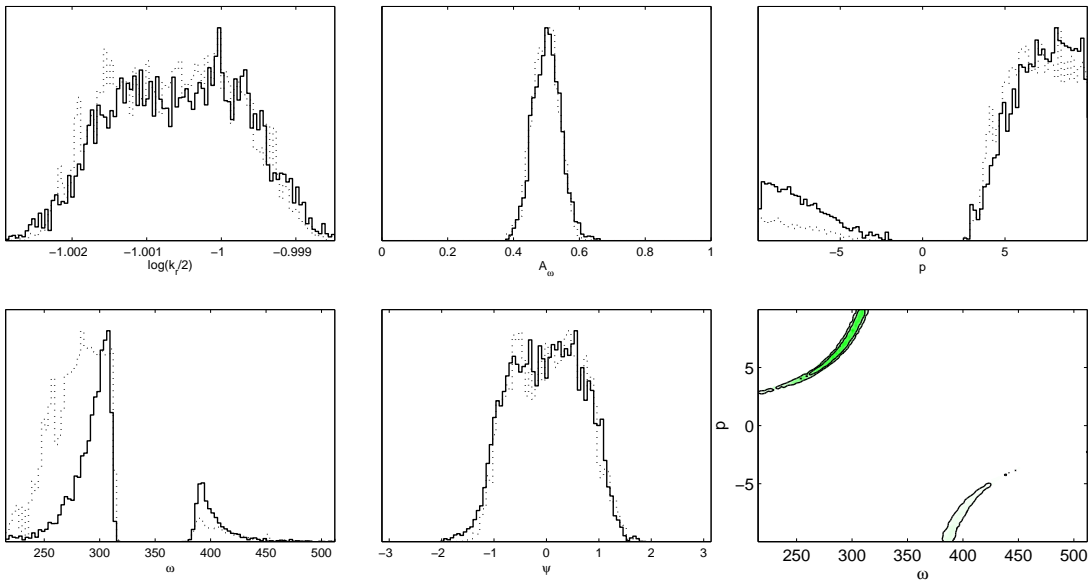}
\caption{Marginalized posteriors for a feature occurring at the small
  scale boundary of the sensitivity zone, $\kf=0.1\,\Mpc^{-1}$. Most of
  the fingerprint parameters are very well reconstructed but not $p$
  and $\omega$ which are strongly degenerated. Their respective one-
  and two-sigma confidence intervals are plotted in the rightmost
  lower frame. This is the result of the noise amputating some part of
  the oscillatory tail. Notice that the inflationary paradigm can
  still be inferred as $|p|>1$.}
\label{fig:aw050_w300_k01}
\end{center}
\end{figure}

At smaller scales, the noise dominates. In
Fig.~\ref{fig:aw050_w300_k01}, we have represented the posteriors
obtained when the feature lies at the small scales boundary of the
sensitivity domain, i.e. for $\kf=0.1\,\Mpc^{-1}$ ($\ellf \simeq
1400$) and for $\fiducial{p}=8$, $\fiducial{\Aw}=0.5$,
$\fiducial{\omega}=300$. There is a net signal detection, the
posterior of $\Aw$ is sharply peaked around the expected fiducial
value ($\Delta \Aw=0.04$), as well for the scale $\kf$ ($\Delta
\kf/\kf = 9 \times 10^{-4}$). However, the frequency probability
distribution appears to be multi-valued as well as the one for the
expansion parameter $p$. In the same figure, we have plotted the one-
and two-sigma confidence intervals associated with the two-dimensional
posterior probability distribution in the plane $(\omega,p)$. Both
parameters end up being strongly correlated as all shaded value
provide a good fit to the CMB resonance pattern. This is
reminiscent with the degeneracy mentioned previously when the
frequency of an inflationary fingerprint is low. Here, compared to the
same feature in the sensitivity zone, part of the oscillatory tail is
actually truncated as becoming of smaller amplitude than the noise
(see Fig.~\ref{fig:cls_aw050}). As a result, there are less measurable
``ticks'' in the CMB and all combination of $p$ and $\omega$ producing
similar oscillations around $\kf$ cannot be distinguished. It is
therefore not surprising that the posteriors obtained here are similar
to the ones of Fig.~\ref{fig:aw020_w100}. Let us notice that the data
still ``see'' that this is an inflationary era -- the oscillations
remain on the right side of $\kf$, independently of the truncation,
and the posterior for $p$ is non-vanishing only when $|p|>1$. Further
out of the sensitivity zone, the fingerprint signals remain
undetectable.

\section{Conclusions and discussions}
\label{sec:conclusion}

We have studied the detectability of the fingerprints induced by
standard clocks in various primordial universe
paradigms. Concentrating on the power spectrum, we have found that
they are detectable in Planck-like data provided they lie in the
sensitivity zone $10^{-2}\,\Mpc^{-1} \lesssim \fiducial{\kf} \lesssim
10^{-1}\,\Mpc^{-1}$. The scale $\kf$, and frequency $\omega$ are the
most sensible parameter to the actual presence of a fingerprint in the
data in most cases. For inflation, although their posteriors exhibit
characteristic signatures for an amplitude as small as
$\fiducial{\Aw}=0.1$ ($10\%$ modulation), we have shown that a proper
reconstruction of the parameter $p$ requires a higher signal
$\fiducial{\Aw} \gtrsim 0.5$, especially at high frequencies ($\omega
\gtrsim 300$). The situation is however improved at lower frequency
($\omega \simeq 100$) as amplitudes $\fiducial{\Aw} \gtrsim 0.2$ still
allow to infer the inflationary paradigm $|p|>1$. Alternatives to
inflation with small $|p| \ll 1$ have been found to deviate
from this rule as very small value of $\fiducial{p}$ end up being more
difficult to detect. As discussed in the previous sections, this is
essentially due to how widespread the observable oscillatory patterns
are. In any case, the high sensitivity zone in $\kf$ is the region to
explore with the soon to be released Planck data.

We have also seen that, for inflation, establishing the bound $|p|>1$
is easier than a full reconstruction of the index $p$. So the main
strength of the standard clocks is to break the leading degeneracy
between the paradigms with $|p|>1$ and $|p|<1$, especially if
they predict degenerated power spectrum and non-Gaussianities in the
absence of the fingerprints. Once this degeneracy is broken, the more
precise value of $p$ could be inferred from other more standard
observables such as the spectral index.

Most likely, hints for a feature could show up while not allowing a
full reconstruction with the CMB power spectrum only. However, these
standard clocks also imprint correlated signals in non-Gaussianities
\cite{Chen:2011zf,Chen:2011tu}. These signals appear as leading order
large non-Gaussianities instead of small corrections, due to the
resonant mechanism~\cite{Chen:2008wn}. The search for such
scale-dependent and non-separable non-Gaussianities is a much more
difficult task however (see e.g. Ref.~\cite{Fergusson:2008ra,
  Meerburg:2009fi, Liguori:2010hx} for possible methods). An efficient
approach would be to search them first in the power spectrum, as we
discuss in this paper. If any candidate signals are found, the
locations of the corresponding non-Gaussianities will be determined,
and the subsequent search for non-Gaussianities would be considerably
narrowed down and provide non-trivial checks.

Finally, let us mention that the matter power spectrum, in the linear
regime, is far less smoothed than the CMB concerning the transfer of
features. However, those resonances are completely washed out by the
galaxy survey window functions~\cite{Martin:2004iv}. A possible future
work, however, may be to discuss how much structure formation may be
affected by features in the non-linear
regime~\cite{FelippeS.Rodrigues:2010xu}.

\acknowledgments

We would like to thank Alan Guth, David Langlois and Anthony Lewis for
helpful discussions. XC is supported by the Stephen Hawking Advanced
Fellowship. CR is partially supported by the Wallonia-Brussels
Federation grant ARC 11/15-040 and the ESA Belgian Federal PRODEX
program $\mathrm{N}^\circ 4000103071$.

\bibliographystyle{JHEP}
\bibliography{fgprints}

\end{document}